\documentclass[prl,twocolumn,a4paper,showpacs,superscriptaddress]{revtex4}

\usepackage{graphicx}
\usepackage{amsmath}

\begin{document}

\title{Critical Behavior of $\textrm{CP}^1$ at $\theta=\pi$, Haldane's 
Conjecture and the Universality Class}

\date{July 18, 2006}
\author{Vicente~Azcoiti}
\affiliation{Departamento de F\'{\i}sica Te\'orica, Universidad de
Zaragoza,
Cl. Pedro Cerbuna 12, E-50009 Zaragoza (Spain)}
\author{Giuseppe~Di Carlo}
\affiliation{INFN, Laboratori Nazionali del Gran Sasso, 67010 Assergi, 
(L'Aquila) (Italy)}
\author{Angelo~Galante}
\affiliation{Dipartimento di Fisica, ITIS "E. Mattei", via S. Rocco, 
66054 Vasto(Italy)}

\begin{abstract}
Using an approach to analyze the $\theta$ dependence 
of systems with a $\theta$-term we recently proposed , the critical 
behavior of $\textrm{CP}^1$ at $\theta=\pi$ is studied. 
We find a region outside 
the strong coupling 
regime where Haldane's conjecture is verified. The critical line however 
does not belong to the universality class of the Wess-Zumino-Novikov-Witten 
model at topological coupling $k=1$ since it shows continuously varying 
critical exponents.

\end{abstract}
\maketitle

Two dimensional $\textrm{CP}^{N-1}$ models are of great interest in high 
energy physics. They are toy models for testing nonperturbative and 
topological properties in a confining, asymptotically free quantum field 
theory and share also with QCD, the theory of the strong interaction between 
particles, a well defined topological charge, possess instanton solutions, 
have $\theta$-vacua and admit a 1/N expansion. 
Understanding the role of the $\theta$ parameter in QCD and its
connection with the ``strong CP problem'' is one of the 
major challenges for high energy theorists \cite{Strongcp}. Unfortunately,
Euclidean lattice gauge theory, our main nonperturbative tool
for QCD studies, has not been able to help us for two reasons: {\it (i)} the 
difficulties in defining the topological charge
operator on the lattice \cite{Top_def} and 
{\it (ii)} the imaginary contribution
to the action coming from the $\theta$ term that prevents the 
applicability of the importance sampling method.
$\textrm{CP}^{N-1}$ models regularized on a lattice have a well defined 
topological charge but share with QCD point {\it (ii)} and this means that,
from the point of view of numerical calculations, the complex
action problem is as severe as in QCD. Indeed most properties of 
$\textrm{CP}^{N-1}$ models have been obtained in the context of the $1/N$ 
expansion \cite{tres, cuatro}, showing a qualitative behavior as a 
function of $\theta$ similar to that of the massive Schwinger model at weak 
coupling \cite{cinco}. The cusp in the free energy density at 
$\theta=\pi$ signals a first order phase transition at this value of $\theta$ 
with spontaneous breaking of CP, in agreement also with the strong 
coupling results \cite{seiberg}. This behavior, which is expected to 
persist for finite values of $N$, at least until $N=4$, is common to most of 
the models the $\theta$ dependence of which is known, and was conjectured 
for non-abelian gauge theories by 't Hooft \cite{HOOFT} in 1981. 
Some time ago, using different arguments and under very general assumptions,
we also argued that a singular behavior in 
$\theta$ is to be expected in systems with a $\theta$-term \cite{PTP}. 
For $N\le 3$ 
dislocations break the asymptotic scaling (continuum limit) \cite{seis} 
and the $\theta$ dependence becomes more intriguing in these cases.

However high energy physics is not the only field where topological structures 
and the $\theta$ term play a fundamental role. In condensed matter physics 
Haldane showed \cite{haldane} that chains of quantum spins with
antiferromagnetic interactions, in the semiclassical limit of large but 
finite $S$, are related to the two-dimensional $O(3)$
nonlinear sigma model at coupling $g=2/[S(S+1)]^{1/2}$ 
and topological term $\theta=0, \pi$ (integer, half-integer spin). 
While integer spin 
chains have a mass gap, half-integer chains should be gapless. 
Haldane conjectured that the $O(3)$ non linear sigma model presents a 
second order phase transition at $\theta=\pi$, keeping its ground state CP 
symmetric. Based on a partial summation of the strong coupling expansion, 
Affleck \cite{affleck} argued that the two-dimensional $O(3)$ non linear 
sigma model has indeed a gapless phase at $\theta=\pi$ and weak coupling.
Furthermore Affleck and Haldane \cite{wznw} also argued that the critical 
theory for generic half-integer spin antiferromagnets is the 
Wess-Zumino-Novikov-Witten model at topological coupling $k=1$, with the 
following scaling law for the mass gap \cite{scaling} near the phase 
transition point:

\begin{equation}
m(\theta) \hskip 0.3cm \alpha \hskip 0.3cm (\pi - \theta)^{2/3} 
ln[(\pi - \theta)^{-1/2}]
\label{scaling}
\end{equation}

In 1995 Bietenholz, Pochinsky and Wiese \cite{bpw} performed a simulation of 
the two-dimensional $O(3)$ non linear sigma model with a $\theta$-term in 
order to verify Haldane's conjecture. They used an efficient Wolff cluster 
algorithm \cite{cluster} on a triangular lattice with a constraint in the 
action which preserves the $O(3)$ symmetry, and simulated the system at a 
fixed value of the coupling (temperature) $g=\infty$ which, due to the 
constraint in the action, should be outside the strong coupling regime. 
By measuring the topological and magnetic susceptibilities and using 
finite size scaling theory, the authors of \cite{bpw} found a second order 
phase transition at $\theta=\pi$ confirming Haldane's conjecture, 
and a finite size scaling in relatively good agreement with the assumption 
that the critical exponents of the phase transition are those of the $k=1$ 
Wess-Zumino-Novikov-Witten model (\ref{scaling}).

As stated before, the sign problem has restricted very much the lattice 
non-perturbative investigations of systems with a $\theta$-term. A few years 
ago we introduced two alternative schemes \cite{monos1, monos2} to analyze 
the $\theta$ dependence in these systems. 
The first approach \cite{monos1} 
was based on a precise determination of the probability distribution 
function of the topological charge whereas the second one \cite{monos2}, 
originally inspired by analytical properties of the
Ising model with an imaginary magnetic field,
is very efficient to determine the critical behavior 
at $\theta=\pi$ in cases in which a continuous transition at this 
value of $\theta$ shows up. 
The two approaches were successfully tested 
in several integrable models and, after this consistency checks, 
applied to the analysis of the $\theta$ 
dependence of the $CP^9$ model. 
Indeed reference \cite{cp9} contains the 
first full reconstruction of the continuum $\theta$ dependence of a 
$\textrm{CP}^{N-1}$ model, showing in particular the spontaneous $CP$ 
symmetry breaking at $\theta=\pi$ in the continuum limit. Here we want to use 
the second of the previously cited approaches \cite{monos2} in order 
to analyze the 
phase structure in $\theta$ of $\textrm{CP}^1$ which, as well known, is 
equivalent to the non linear $O(3)$ sigma model. 
For this reason in the following we will briefly summarize the main
algorithm steps. 

We have adopted for
the action the standard ``auxiliary U(1) field'' formulation
\begin{equation}
S_g = 2\beta\sum_{n,\mu}(\bar{z}_{n+\mu}z_n U_{n,\mu} +
\bar{z}_n z_{n+\mu} \bar{U}_{n,\mu}-2)
\label{action}
\end{equation}
where $z_n$ is a $2$-component complex scalar field that satisfies
$\bar{z}_n z_n = 1$ and $U_{n,\mu}$ is a $U(1)$ ``gauge field''.
The topological charge operator is defined directly from the $U(1)$ field:
\begin{equation}
S_\theta =
i \frac{\theta}{2\pi}\sum_p\log(U_p)
\label{topocharge}
\end{equation}
where $U_p$ is the product of the $U(1)$ field around the
plaquette and $-\pi < \log(U_p) \le \pi$.

Our numerical approach uses as input the results from numerical simulations 
of $\textrm{CP}^1$ at imaginary values of $\theta= -ih$ ($h$ real) 
\cite{monos2}. The action to simulate $S=S_g + S_{\theta=-ih}$ is then 
real and local. This implies that the main computational cost
is practically equivalent to standard simulations
of $\textrm{CP}^1$ model at $\theta=0$.
This is important since it allows to perform large volume simulations. 

We use $x(h)$, the topological charge density, and the
variable $z=\cosh\frac{h}{2}$ to define the quantity 
$y(z)=x(h)/\tanh\frac{h}{2}$. 
Using the transformation $y_\lambda(z)=y(e^{\frac{\lambda}{2}}z)$
and plotting $y_\lambda/y$ against $y$  
one gets typically a smooth function
for small $y$ \cite{monos2}. Hence 
we can expect a simple extrapolation to $y\to 0$ ({\it i.e.} in
the region corresponding to real $\theta$) to be reliable and
thus obtain $x(\theta)$ (of course the result has to be independent of the
specific value of $\lambda$). 
In the same spirit the effective exponent
$\gamma_\lambda=\frac{2}{\lambda}\log(y_\lambda/y)$ as a function of
$y$ will give the dominant power of $y(z)$ as a function of $z$ near
$z=0$ or, equivalently, the behavior of $x(\theta)$ for $\theta\to\pi$.
If $\gamma_\lambda(y\to 0)=1$ $\textrm{CP}$ symmetry is spontaneously
broken at $\theta=\pi$, values between 1 and 2 indicate a second
order phase transition with a divergent susceptibility and so on 
\cite{monos2}. Our approach assumes implicitly that the theory has not 
phase transitions at $\theta<\pi$ and in particular that $CP$ symmetry 
is realized at $\theta=0$, the last being a necessary condition for the theory 
to be well defined at $\theta\ne 0$ \cite{pct}.

Fig. 1 shows the goodness of our method when applied to two models, the 
analytical solution of which is known \cite{monos2}. The first one is the 
one-dimensional Ising model within an imaginary magnetic field, 
which breaks spontaneously $CP$ at $\theta=\pi$. The second one is a toy 
model with only one effective degree of freedom. This model, 
which resembles very much the free instanton gas model, has the partition 
function ${\cal Z}_V(\theta)=\left( 1+A \cos\theta\right)^V$ for 
$V$ degrees of freedom. We have plotted in Fig. 1 the order parameter 
$x(\theta)$ against $\theta$ for both models. The lines stand for the exact 
analytical solution whereas the symbols correspond to the values extracted 
from our approach.

\begin{figure}[t!]
\centerline{\includegraphics*[width=2in,angle=270]{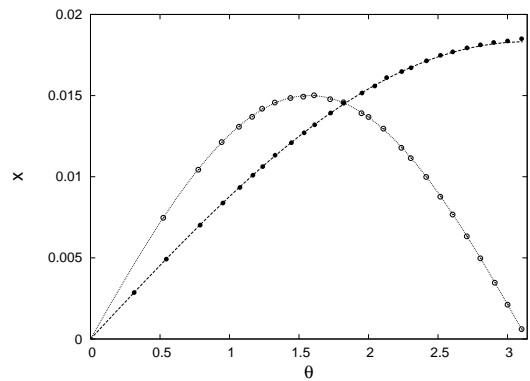}}
\caption{Order parameter in Ising model (full symbols) and 
a toy model discussed in the text 
(open symbols); the lines are analytical results
}
\label{fig 1}
\end{figure}

Let us resume our results for $\textrm{CP}^1$. This model is equivalent, 
at $\beta=0$ (strong coupling), to two-dimensional compact $QED$ with 
topological charge. $CP$ symmetry is therefore spontaneously broken at 
$\theta=\pi$ in this limit. This result
remains qualitatively unchanged when 
moving to larger $\beta$ values, until $\beta\sim 0.5$. For $\beta > 0.5$ 
$CP$ symmetry is recovered at $\theta=\pi$ and this result seems to hold 
for any $\beta$. For $\beta$ values between $0.5$ and $1.5$ the order 
parameter $x(\theta)$ vanishes at $\theta=\pi$ as 
$(\pi-\theta)^{\epsilon(\beta)}$ with $\epsilon(\beta)$ varying continuously 
between $0$ and $1$. This implies a divergent topological susceptibility 
$\chi_t= {{dx(\theta)}\over{d\theta}}$ at $\theta=\pi$ 
for $0.5< \beta<1.5$ and 
we get therefore a line of second order phase transition points with 
continuously varying critical exponents. 
For $\beta>1.5$ we find $\epsilon(\beta)=1$. 
Numerical evidence of these results is given in figures 2-5. 

The details of the simulations are: we used lattices $40^2$ 
for $\beta \leq 1.0$, $100^2$ for $1.0 \leq \beta \leq 1.5$ and 
$200^2$ for $\beta \geq 1.5$ checking absence of finite volume effects;
this choice has been taken looking at existing data for the correlation
lenght at $\theta=0$ \cite{picp1} ($\xi < 3$ for $\beta \leq 1.0$, 
$3<\xi < 20$ for $1.0<\beta < 1.5$). For each value of $\beta$ we have
performed simulations for 25-40 values of $h$ with statistics exceeding
500000 MC iterations each in order to evaluate $x(h)$.
 
In Fig. 2 we report the order parameter as a function of $\theta$
for $\beta=1.0$, hence in the region where $0< \epsilon <1$; 
in Fig. 3 the same data are plotted in a log-log scale and, to
give the feeling of the accuracy in the determination of the critical
exponent, a fit
with the function $x(\theta)=const+a(\pi-\theta)^{\epsilon(\beta)}$ 
is also shown. The fitted parameters are: const=0, a=0.441(1) and 
$\epsilon(1.0)=0.736(1)$.
Equivalent graphs have been obtained for the other values of $\beta$ in the
range 0.6 - 1.5 
while, for $\beta \leq 0.5$, 
we have a different behaviour of
the order parameter, approaching a non zero value at $\theta=\pi$.
For the two larger values of $\beta$, namely $1.5$ and $1.6$, we obtain
$\epsilon=1$ and an order parameter indistinguishable from a 
$\sin(\theta)$ function between $\theta=0$ and $\theta = \pi$. 

\begin{figure}[t!]
\centerline{\includegraphics*[width=2in,angle=270]{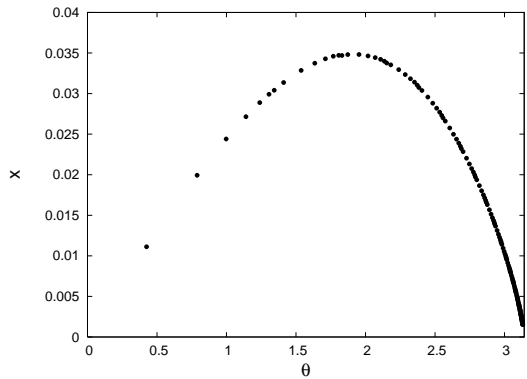}}
\caption{ Order parameter for $CP^1$ model; $\beta=1.0$ , L=100. Statistical
errors are of the order of the symbol size.}
\label{fig 2}
\end{figure}

\begin{figure}[t!]
\centerline{\includegraphics*[width=2in,angle=270]{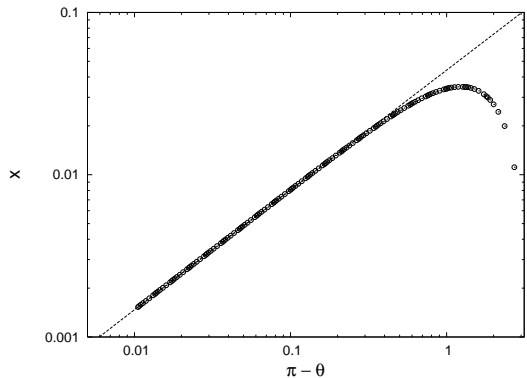}}
\caption{ The same data of Fig. 2 in log-log plot; the line is the fit
with the funcion $const+a(\pi-\theta)^{\epsilon(\beta)}$ }
\label{fig 3}
\end{figure}

Starting from the same raw data (the simulations at imaginary $\theta$)
we can do a consistency check performing a different analysis using 
the effective exponent $\gamma_\lambda$. 
We are interested in the effective exponent in the $y\to 0$ limit
and, in Fig. 4, we can see the $\gamma_\lambda$ extrapolation 
for $\beta=1.0$ ($L=100$ data, lower curve) and for $\beta=1.6$ ($L=200$
data, upper curve). For the $\beta=1.0$ case extrapolations
obtained using two different fitting functions, namely $\gamma_\lambda(y)=
\exp(a + b y + c y^2)$ and $ a + b y + c y^2 $ are shown; the first
function is able to slightly better reproduce  the data in the full $\beta$ 
range with respect to the simple polynomyal form.
The small difference in the extrapolated values does not change qualitatively
the result. 

For the $\beta=1.6$ data there is essentially no need for extrapolation,
being very clear that the results point to the value $\gamma=2$ in the $y=0$ 
limit. Data shown in the figure are those relative to $\lambda=0.5$
but exactly the same conclusions can be reached using other values of
$\lambda$. 
It is important to notice how, moving to the region of smaller $\beta$, 
the same fitting functions continue to describe very well the data and 
have been used to extrapolate the effective exponent at zero $y$.

\begin{figure}[t!]
\centerline{\includegraphics*[width=2in,angle=270]{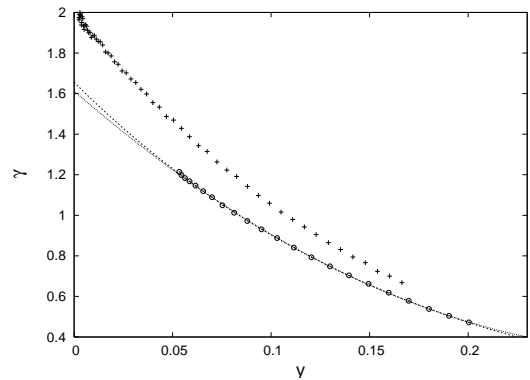}}
\caption{ Effective exponent $\gamma_\lambda$ for $\beta=1.0$, 
L=100 (circles) and $\beta=1.6$ , L=200 (crosses) and two different 
extrapolations of $\beta=1.0$ data.}
\label{fig 4}
\end{figure}

Fig. 5 resumes all results for exponents $\epsilon$ and $\gamma$ as
a function of $\beta$: it is clear from this figure 
that, considering the fate of $CP$ symmetry at $\theta=\pi$,
the system passes from a phase in which the symmetry is broken
($\beta < 0.5$) to a symmetric phase ($0.5< \beta < 1.5$) where a 
second order phase transition with varing critical exponents is present,
ending with a region in which $CP$ symmetry is still satisfied but 
without a divergent correlation lenght. 

\begin{figure}[t!]
\centerline{\includegraphics*[width=2in,angle=270]{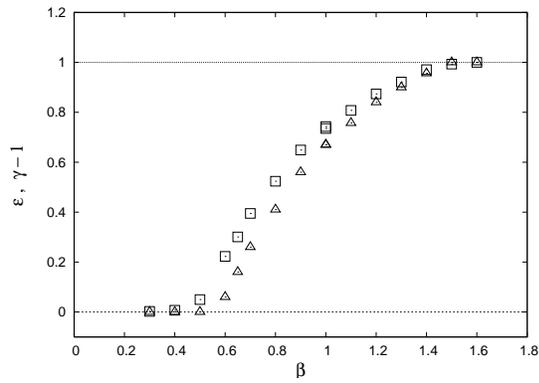}}
\caption{Critical exponent $\epsilon$ (squares) from fits to the
order parameter behaviour and effective exponent $\gamma_\lambda$
(actually $\gamma_\lambda -1$ 
in order to be directly comparable with $\epsilon$ , 
triangles).}
\label{fig 5}
\end{figure}

The location of the point where the system becomes 
$CP$ symmetric at $\theta=\pi$ can not be obtained from our data with
a high accuracy since different analysis gives slightly different results
but the existence of such a point is evident from both analysis presented 
here.

Nevertheless, in our opinion, the important physical fact is the 
existence of a line of second order transition points with 
continuously varing critical exponents. 
Using the hyperscaling hypothesis we can relate 
$\epsilon(\beta)$ with the critical exponent $\nu$ for the mass gap. 
Indeed, following Kadanoff, one can assume that the singular part of the 
free energy 
density near the phase transition points is proportional to $\xi^{-2}$, 
$\xi$ being the inverse mass gap, and this implies that 
$m(\theta)\sim (\pi-\theta)^\nu$. 
As a consequence simple 
algebra gives the relation $\epsilon = 2\nu -1$.

In conclusion we have analyzed the critical
behavior of $\textrm{CP}^1$ at $\theta=\pi$ and found a region in the coupling 
$\beta$ where Haldane's conjecture is verified. The critical line shows 
continuously varying critical exponents and does not belong therefore 
to the universality class of the Wess-Zumino-Novikov-Witten
model at topological coupling $k=1$. From the point of view of Quantum Field 
Theory these new critical points open the possibility for new nongaussian 
fixed points with anomalous dimensions, where a non trivial continuum limit 
could be defined.

\end{document}